# Fabrication of a 600V/20A 4H-SiC Schottky Barrier Diode

**In-Ho Kang, Sang-Cheol Kim, Jung-Hyeon Moon, Wook Bahng, and Nam-Kyun Kim**

*Power Ssemiconductor Research Center, Korea Electrotechnology Research Institute (KERI),*

*Changwon 642-120, Korea*

600V/20A 4H-SiC Schottky barrier diodes (SBD) were fabricated to investigate the effect of key processing steps, especially before and after a formation of Schottky contact, on the electrical performances of SBD and on a long-term reliability. The results show that 4H-SiC SBDs, subjected to a hydrogen-ambient annealing at 470$^o$C for 10min and sacrificial treatment right after ion activation, exhibited low forward voltage drop ($V_F$) at a rated current of 20A, higher blocking voltage of 800V, and very short reverse recovery time of 17.5ns. Despite a harsh reverse bias condition and temperature, a long-term reliability test shows that changes in forward voltage drop and reverse leakage current ($I_R$) were 0.7% and 8.9% and that blocking voltage was enhanced. This is attributed to a stabilized interface between passivation layer and SiC that is caused by aging.

# I. INTRODUCTION

4H-SiC is a promising material for efficient power devices that are able to block a higher voltage even with low power loss, to enable advanced future power networks including smart grids and high-voltage DC power transmission, due to the superior material properties of 4H-SiC [1]. A 4H-SiC Schottky barrier diode (SBD) among various semiconductor devices was firstly commercialized and has been widely used in the power electronic systems due to its low conduction loss and nearly-zero reverse recovery time. In spite of many reports on the fabrication of 4H-SiC SBDs having a low current rating (< 1 A) and prevailing commercial products, the influence of multiple processing steps on the electrical performances of the 4H-SiC SBDs having a large area that enables higher current driving capability has been still elusive to date.

In this paper, we introduce the whole processing steps to fabricate a 600V/20A 4H-SiC SBD and investigate the effect of key processing steps, especially before and after a formation of Schottky contact, on the electrical performances of SBD and on a long-term reliability.

# II. EXPERIMENT

Figure 1 shows a schematic cross-section of a 4H-SiC Schottky barrier diode (SBD). The device structure is composed of passivation layer, Schottky and pad metals, a 0.6-μm-thick $p^+$ layer, and an 8-μm-thick n- drift layer on an $n^+$ 4H-SiC substrate. The $p^+$ layers serve simultaneously as a p-grid that forms a junction barrier Schottky (JBS) and has immunity to surface defects and as the guard ring (G/R) and field limiting ring (FLR) that serve as an edge termination to reduce the field crowding near the edge of the Schottky junction, thereby leading to a higher breakdown voltage. Generally two distinct masks are used to form the p-grid for JBS structure and the edge termination. In this case, a single implantation was used, leading to reduced number of mask and lower production cost.

The SBD was fabricated on a commercially-available 3" 4H-SiC epiwafer. The $p^+$ layer, having a doping concentration of $5 \times 10^{18}$ cm$^{-3}$ was implemented by using multiple Al-ion implantations to form a box-like doping profile. The ion implantation process was done at an elevated temperature of 650°C in order to reduce the lattice damage during implantation. Subsequently, all implanted regions were activated for a lattice healing and an electrical activation of dopant ions. The activation process was executed at 1700°C for 60 min under an Ar atmosphere by using a carbon cap layer to protect the surface. The carbon cap layer was removed by oxidation at 1150°C for 2 hours under a dry $O_2$ atmosphere, followed by an oxide etch using buffered oxide etch (BOE) for 10 min. A 5-hour sacrificial oxidation in the same condition was also done to reduce the surface leakage current caused by a C-rich surface layer due to preferential Si sublimation during high-temperature activation. Additional oxidation for 5 hours was done to form the first passivation layer. A 200-nm-thick Ni-V layer was deposited by sputtering and annealed at 950°C for 90 sec to form back-side ohmic contact. To investigate the effect of Schottky metals and post-treatment after deposition of them on the electrical performances of SBD, a 300-nm-thick various metals as a Schottky metal and a 4-um-thick Al as a pad metal were deposited after opening the oxide. Subsequent annealing at various temperatures under a hydrogen atmosphere was done. Finally the SBDs were passivated by a polyimide (PI) at 950°C for 1 hour under $N_2$ atmosphere. SBDs were electrically characterized using an HP4156 semiconductor parameter analyser and a Tek371A curve tracer to measure forward I-V characteristics and an in-house current-voltage characterizer to measure reverse I-V characteristics. To perform high-temperature reverse reliability test, an in-house setup was employed..

## III. RESULTS AND DISCUSSION

It is well known that higher Schottky barrier height (SBH) reduces reverse leakage current (JRP), leading to higher blocking voltage, while increases forward voltage drop (VF), leading to higher conduction loss [2]:

$$V_F = \frac{nkT}{q} \ln\left(\frac{J_F}{A^{**}T^2}\right) + n\phi_B + R_{on}J_F \tag{1}$$

and

$$J_{RP} = A^{**}T^2 \exp\left(-\frac{\phi_B}{kT}\right) \exp\left(-\frac{\Delta\phi_B}{kT}\right), \tag{2}$$

where k is Boltzmann's constant, q is the electron charge, T is the temperature, n is the ideality factor, $\phi_B$ and $\Delta\phi_B$ are SBH and a reduced SBH caused by a Schottky barrier height lowering, $J_F$ is the forward current density at $V_F$, $A^{**}$ is Richardson's constant, and $R_{on}$ is the drift region resistance. Therefore, Schottky barrier height must be optimized to simultaneously obtain lower $V_F$ and higher blocking voltage (or lower leakage current). Table 1 shows comparison of key parameters, including SBH, ideality factor, $V_F$, and leakage current, for various Schottky metals and annealing conditions. SBH for SBD with Ni/Al as Schottky metal decreased with increasing annealing temperature. It was reported that Ni/Al could produce moderate ohmic contacts for 4H-SiC at relatively low annealing temperature of 800°C due to formation of $Ni_2Si$ and $NiAl_3$ compounds [3]. In this work, more reduction in SBH for SBD with Ni/Al was observed than that for SBD with Ni/TiW/Al, where TiW serves as a diffusion barrier for Al. As a result, it is easily expected that a reaction with aluminum strongly affected reduction in SBH through formation of $NiAl_3$, although formation of $Ni_2Si$ could lower SBH. However SBH for SBD with Ti/Al as Schottky metal showed highest value at 470°C and decreased with increasing annealing temperature. According to the previous work [4], SBH for Ti/Al could increase with annealing temperature from 500°C to 750°C mainly due to a Fermi-level pinning effect caused by deep trap level at interface between Ti and 4H-SiC and to a formation of TiC. Therefore, at annealing temperature of 470°C, SBH increased as aforementioned, while decreased due to diffusion of Al above 500°C. This is also strongly supported by a monotonic increase in SBH for SBD with Ti/TiW/Al with annealing temperature, where TiW serves as the diffusion barrier.

In the next step, to select best Schottky barrier height that meets the above constraints, the simulation results of Dahlquist were taken into account [5]. The results show that SBH of 1.3eV, 8-μm epilayer, and doping concentration of $2\times10^{16}\text{cm}^{-3}$ were best choice to achieve trade-off between forward voltage drop and leakage current, if 80% critical field of 4H-SiC was considered for process margin. In this work, SBH of 1.124eV was adopted to reduce forward voltage drop, while thicker epilayer of 8μm and lower doping concentration of $1\times10^{16}\text{cm}^{-3}$ were utilized to reduce leakage current.

Figure 2 shows (a) forward I-V characteristics and (b) re-verse I-V characteristics of 600V/20A SBD. The chip size of SBD is $3.6\times3.6\text{mm}^2$. The forward voltage drop ($V_F$) is 1.77V measured at $I_A$=20A. The equivalent $R_{on}$ is 5mΩ·cm², which is a little bit larger than that of commercially available SBD. This is attributed to higher area ratio of p-grid over pure Schottky, which must be further optimized to achieve lower $R_{on}$. However thermal reliability is expected to be improved because the current density is 154A/cm² and low enough for air-cooling. The reverse leakage current is 20μA measured at $V_A$=-600V. This is a little bit low even for large chip area. We believe that a surface treatment, composed of the 5-hour sacrificial oxidation and oxide removal, eliminated the surface damage generated during high temperature activation, leading to removal of surface leakage path and to lower leakage current [6]. Figure 3 shows the distributions (a) of SBH and (b) of reverse leakage current measured at VA=-600V and blocking voltage measured at $I_A$=-200μA for 100 SBDs. In spite of surface treatment, reverse leakage current and blocking voltage are widely spread, while SBHs are relatively concentrated at 1.2eV. As mentioned before, SBH determines leakage current and blocking voltage of ideal Schottky diode. Hence the distributions of leakage current and blocking voltage should be similar with that of SBH. Recently, it was reported that the leakage current of SBD increased with treading dislocation density despite no surface crystal defects within the active area and a normal ideality factor [7]. In other words, the SBDs having large area and even good ideality factor may exhibit unexpectedly large leakage current due to an increasing number of treading dislocations with increasing area.

Figure 4 shows the reverse recovery characteristics of 600V/20A SBD measured at di/dt=200A/μs. A peak reverse current and a reverse recovery time are 2A and 17.6ns, which are very low because SBD is a majority carrier device.

To check long-term reliabilities for the forward voltage drop, the reverse leakage current, and the blocking voltage, the I-V characteristics of SBDs having the same initial values had been measured at 175°C for 1000 hours. Figure 5(a) shows the forward voltage drop measured at $I_A$=10A. The mean values are within 1.44V to 1.45V and the standard deviation value is 0.03V. The long-term deviation of the forward voltage drop is very small even at high temperature. Generally SBDs subject to long-term reliability test conducted at high temperature undergo an aging effect, leading to large deviation in the aforementioned key parameters, especially at an initial stage, because of a weaker metal-semiconductor junction than a p-n junction. Therefore we think that the reason for such a small deviation in forward voltage drop is that the annealing was carried out at 470°C under $H_2$ atmosphere for 10 min, which resulted in stabilization of the metal-semiconductor junction. Figure 5(b) shows the reverse leakage current measured at $V_A$=-600V. The mean values are within 92µA to 101µA and the standard deviation values are within 46µA to 50µA. In other words, the mean reverse leakage current has been decreased by 8.9% for 1000 hours. Figure 5(c) shows the blocking voltage measured at $I_A$=-200µA. The mean values are within 619V to 632V and the standard deviation values are within 31V to 47V. Note that the mean blocking voltage increases and the mean reverse decreases as the aging time increases. Generally the blocking voltage has very strong dependence on the leakage current. The leakage current ($J_R$) of JBS diode can be expressed as the follow:

$$J_R = q \frac{\sqrt{D_n}}{\tau_n} \frac{n_i^2}{N_A} + \frac{q n_i W}{\tau_p} + J_{RP} + \frac{V}{\rho \, l_{eff}} \tag{3}$$

where $D_n$ is the electron diffusion constant, $\tau_n$ and $\tau_p$ are electron and hole life time, $n_i$ is the intrinsic carrier density, $N_A$ is the acceptor concentration, W is the depletion width, V is the reverse bias, $\rho$ is the resistivity of SiC, and $l_{eff}$ is an effective periphery length of metal contact.

The first term of equation 3 describes the leakage current component for the p-n junction of JBS diode, which cannot be affected during the long-term aging because the p-n junction is the most stable.

The second term of equation 3 describes the leakage current caused by traps in the depletion region, which cannot be also affected at such a low test temperature. The third term expresses the leakage current for Schottky junction, which was found to be not affected as mentioned earlier. The last term is concerned with surface leakage current which flows along the periphery of metal contact. The surface state underneath the passivation layer near the edge termination, where electric field is crowded, has strong impact on the reverse I-V characteristics of power device, because the surface or interface charges change the electric field distribution, leading to a premature breakdown and the damaged surface serves as a leakage current path [6, 8]. In a recent report, the power device which has undergone a high reverse bias-stress exhibited the improved reverse characteristics including a reduced leakage current and an increased blocking voltage, by applying a high reverse bias stress, generating joule heating stress, and consequently removing the damaged surface. In the similar way, we believe that the high-temperature, high-reverse bias stress stabilized the interface between passivation layer and SiC by removing the damaged surface and thus induced a lower leakage current and a higher blocking voltage after long-term reliability test.

## IV. CONCLUSION

The effects of key processes on the I-V and reverse recovery characteristics and the long-term reliability of 600V/20A 4H-SiC Schottky barrier diodes have been investigated. To achieve optimum process conditions for a good Schottky contact and for the improved reverse characteristics, Schottky barrier engineering for various Schottky metals and annealing conditions was conducted. The SBD, which underwent a annealing at $470^{o}C$ under $H_2$ atmosphere for 10 min after deposition of Schottky metal and a sacrificial oxidation after high temperature activation, has exhibited lower forward voltage drop, lower leakage current, higher blocking voltage, and good long-term reliability.

## ACKNOWLEDGMENTS

This research was supported by a Korea Electrotechnology Research Institute (KERI) grant.

# TABLES

Table. 1 Comparision of key parameters for various Schottky metals and annealing conditions

| Schottky metals and annealing conditions | | Schottky barrier height (eV) | Ideality factor | $V_F$ (@1mA/cm$^2$) | Leakage current @ 50V (uA) |
|---|---|---|---|---|---|
| Ti(300n) /Al(4um) | Before annealing | 0.96 | 1.063 | 0.365 | 1 |
| | 470°C, 10min | 1.124 | 1.189 | 0.64 | 0.05 |
| | 500°C, 20min | 1.082 | 1.085 | 0.67 | 0.5 |
| | 550°C, 20min | 1.029 | 1.136 | 0.49 | 5.7 |
| Ni(300nm) /Al(4um) | Before annealing | 1.524 | 1.15 | 0.955 | 0.0001 |
| | 470°C, 10min | 0.725 | 1.327 | 0.170 | 194 |
| | 500°C, 20min | 0.869 | 1.015 | 0.170 | 28 |
| | 550°C, 20min | 0.902 | 1.411 | 0.310 | 7.6 |
| Ni(100nm) /TiW(200nm) /Al(4um) | Before annealing | 1.645 | 1.104 | 1.040 | 0.0038 |
| | 470°C, 10min | 1.633 | 1.007 | 1.090 | 0.00042 |
| | 500°C, 20min | 1.458 | 1.496 | 0.940 | 0.00055 |
| | 550°C, 20min | 1.395 | 1.209 | 0.825 | 0.00137 |
| Ti(100nm) /TiW(200nm) /Al(4um) | Before annealing | 0.875 | 1.012 | 0.280 | 0.6 |
| | 470°C, 10min | 0.939 | 1.028 | 0.340 | 0.8 |
| | 500°C, 20min | 0.939 | 1.029 | 0.35 | 2.4 |
| | 550°C, 20min | 0.975 | 1.155 | 0.410 | 0.4 |

**FIGURES**

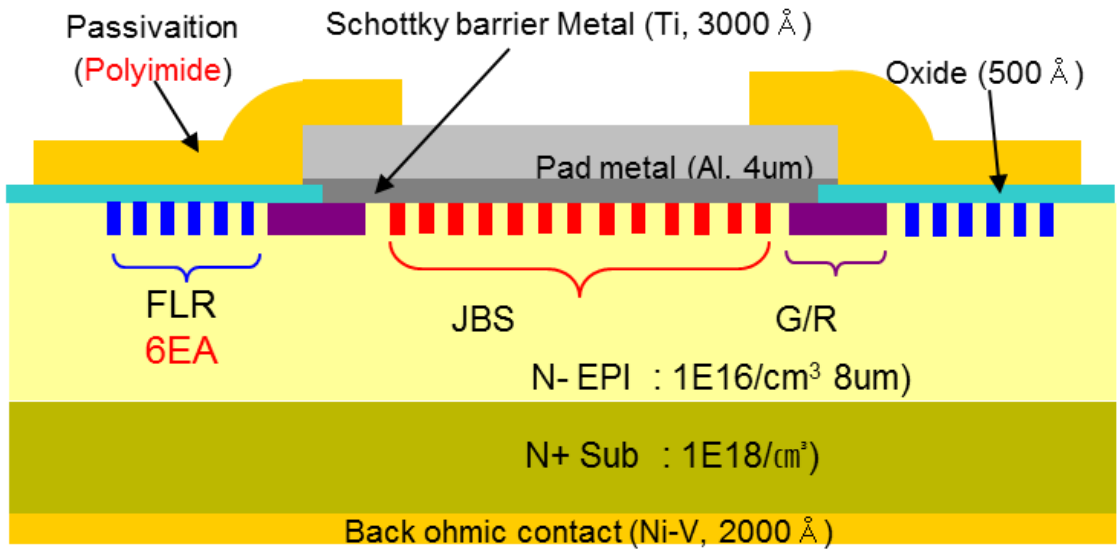

Fig. 1. Cross-sectional view of 4H-SiC Junction Barrier Schottky (JBS) diode.

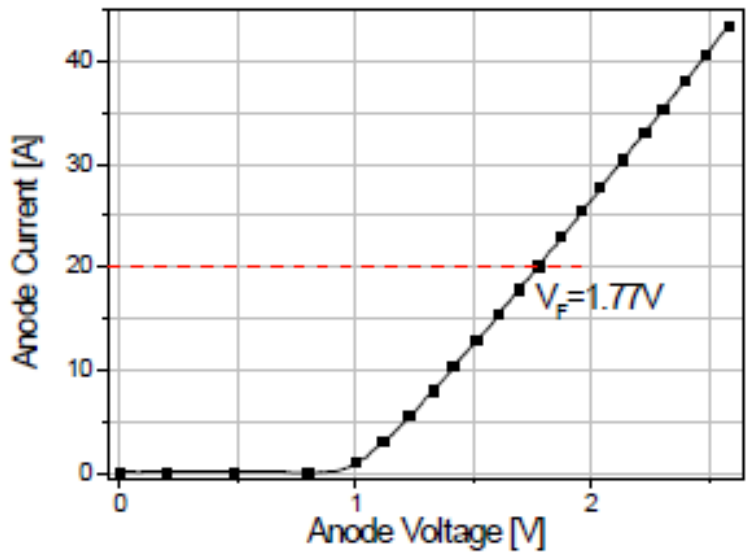

(a)

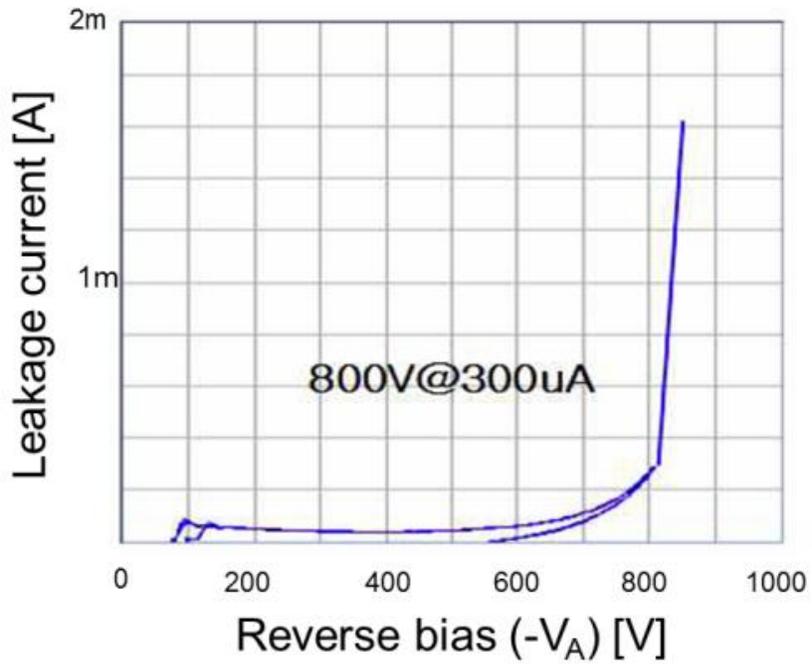

(b)

Fig. 2. Electrical performances of SBD: (a) forward I-V curve, (b) reverse I-V curve, and (c) reverse recovery

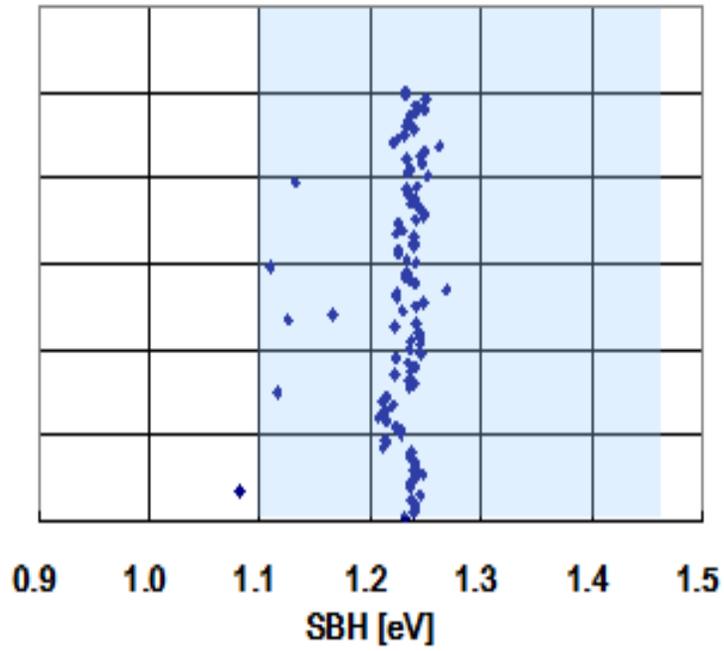

(a)

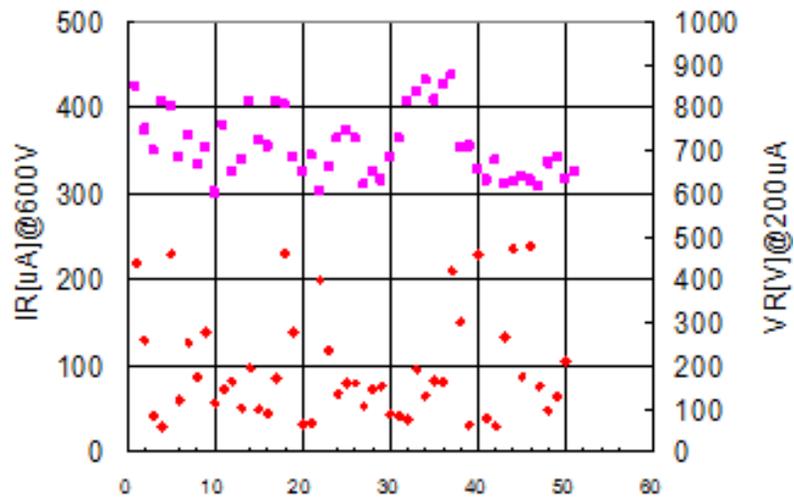

(b)

Fig. 3. (a) distribution of SBH and (b) distribution of reverse leakage current (small red dot) and blocking voltage (large pink dot) for 100 SBDs.

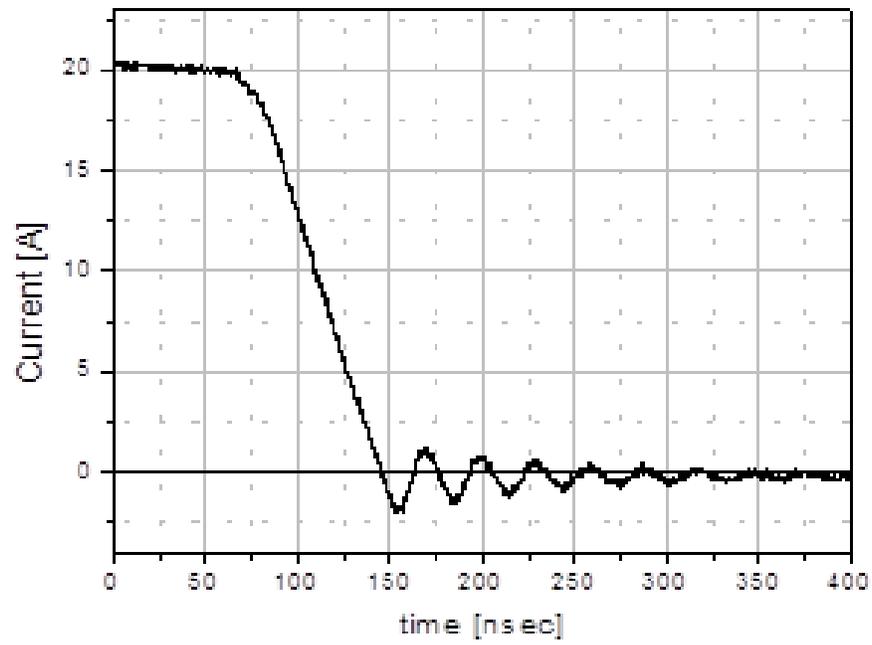

Fig. 4. Reverse recovery characteristics of 600V/20A SBD.

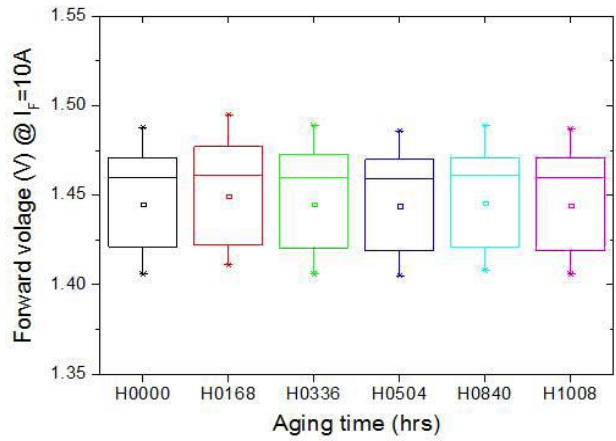

(a)

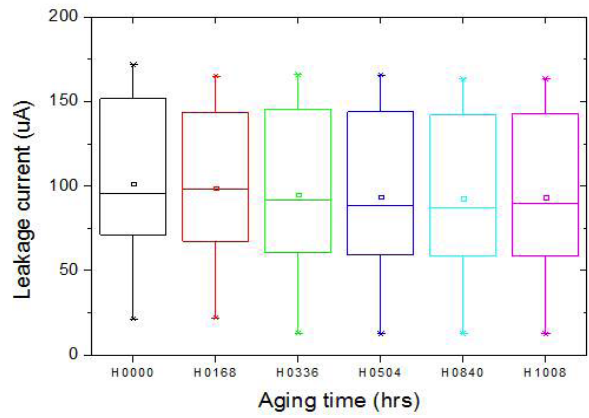

(b)

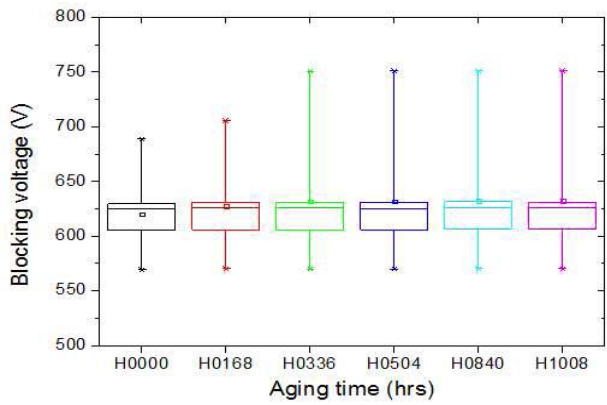

(c)

Fig. 5. long-term reliability for (a) forward voltage drop ($V_F$), (b) reverse leakage current, and (c) blocking voltage.

14